\documentclass[useAMS,usenatbib]{mn2e}
\usepackage{times}
\usepackage{epsfig}

\newcommand{\etal}
	{et al.}
\newcommand{\eg}
	{e.g.}

\newcommand{\bj}
        {b_{\rm J}}

\title[Spectral parameterisation in the 2dFGRS as a diagnostic of star formation history] 
{Galaxy spectral parameterisation in the 2dF Galaxy Redshift Survey}

\author[D.S. Madgwick \etal]    
       {Darren S.\ Madgwick$^{1,2}$\thanks{
                E-mail: dsm@astron.berkeley.edu}, 
        Rachel Somerville$^{3}$,
        Ofer Lahav$^{2}$ \& 
        Richard Ellis$^{4}$ \\ 
        $^1$Department of Astronomy, University of California,
        Berkeley, CA 94720, USA \\
        $^2$Institute of Astronomy, Madingley Road, Cambridge
        CB3 0HA, U.K. \\
        $^3$Department of Astronomy, University of Michigan,
        Ann Arbor, MI 48109-1090, USA \\
        $^4$Department of Astronomy, California Institute of
       Technology, Pasadena, CA 91125, USA \\}
\begin{document}

\date{
Accepted. 
Received; In original form; 11 September 2002}

\pagerange{\pageref{firstpage}--\pageref{lastpage}}
\pubyear{2001}

\maketitle

\label{firstpage}

\begin{abstract}
We investigate the physical significance of a new spectral
parameter, $\eta$.  This parameter was defined based on a Principal 
Component Analysis of the 2dF Galaxy
Redshift Survey, to retain astrophysical information while minimising
the effect of measurement uncertainties.
We find that
while $\eta$ is correlated with morphological type, there is a large
scatter in this relationship. A tighter empirical relationship is
found between $\eta$ and the equivalent width of the H$\alpha$ line,
suggesting a connection with the star formation rate. We
pursue this connection using spectral synthesis models. Using models
in which the star formation history is parameterised in terms of an
exponentially decreasing function of time, we find that there is a
tight correlation between $\eta$ and the ratio of the present to the
past-averaged rate of star-formation, often known as the ``birthrate''
parameter $b$. This correlation also holds in models with much more
complicated star formation histories, generated by a semi-analytic
model of galaxy formation based upon the hierarchical formation
scenario.  
There are two possible causes for the tight correlations we find, 
between $\eta$ and $b$, in those galaxies with the most complex star 
formation histories;
Firstly, the spectra themselves may be 
degenerate to the actual long-term star formation history of each galaxy
in the optical wavelength range probed by the 2dFGRS.  Secondly, 
the birthrate parameter
$b$ may represent a physically fundamental quality of galaxy halos -- their
over-density relative to the background density -- such that small-$b$
galaxies form in high peaks (which collapse early) while large-$b$
galaxies represent lower peaks, which collapse later. We conclude that
the tight connection with $b$ makes $\eta$ a physically
meaningful, as well as convenient and robust, statistic for galaxy 
parameterisation and classification.
\end{abstract}


\section{Introduction}
\label{section:intro}

Galaxy redshift surveys are now probing the galaxy distribution of the
local Universe more accurately than ever before, and in so doing they
are establishing many fundamental properties of the galaxy population
and its large-scale structure.  The 2dF Galaxy Redshift Survey
(2dFGRS) is one such ambitious project, conceived with the aim of
mapping the galaxy distribution to an
extinction corrected 
magnitude limit of $b_{\rm J}\sim19.5$.  This
particular survey is now essentially complete with approximately
230,000 galaxy spectra having already been acquired, and has already
started to yield significant results (e.g. Percival \etal\ 2001;
Madgwick \etal\ 2002 and others).  In addition to this survey, the
Sloan Digital Sky Survey (Strauss \etal, 2002) is also underway, 
and once complete will obtain up to 1,000,000
individual galaxy spectra.
 
Apart from the main scientific goals of quantifying the large-scale
structure of the Universe, one of the most significant contributions
of galaxy redshift surveys is to our understanding of the galaxy
population itself, through the information about galaxy properties
contained in the observed spectra.  Having a data set of 230,000
galaxy spectra --- as in the case of the 2dFGRS --- allows us to test
the validity of galaxy formation and evolution scenarios with
unprecedented accuracy.  However, the sheer size of the data set
presents its own unique problems.  Clearly, in order to make the
spectral data set more `digestible', some form of data compression is
necessary. Familiar statistics such as equivalent width measurements,
morphological types, and broad-band colours are really just
compression techniques in some sense.  These quantities can be
compared with theoretical predictions and simulations, and hence can
set constraints on scenarios for galaxy formation and biasing.

The approach that has been adopted in Madgwick et al. (2002)
is to define a spectral indicator, $\eta$, based on Principal
Component Analysis (PCA).
Because of instrumental limitations, the flux calibration of the 2dF
spectra is unreliable, and hence robust measurements of the shape of
the spectral continua are not possible. For this reason, the $\eta$
statistic was designed to (in a loose sense) preserve the most
discriminatory information in the spectra, while being robust to the
instrumental uncertainties. Effectively, the parameter $\eta$ measures
the strength of absorption and emission lines, while remaining
insensitive to the slope of the continuum (or broad band colour). 

The motivation of the definition of the $\eta$ statistic was in part
pragmatic. The goal of this paper is to address the fundamental
question of the \emph{physical significance} of this parameter, in
terms of how it relates to the physical properties of galaxies. A
substantial amount of work has been carried out in the past on
relating the observed spectra of galaxies to the physical processes
occurring in them (see e.g Kennicutt, Tamblyn \& Congdon 1994; Ronen,
Ar\'agon-Salamanca \& Lahav 1999; Carter, Fabricant, Geller \& Kurtz
2001 and references therein), and the advent of a large and
uniform survey such as the 2dFGRS will result in many more advances in
this field.  However, for the present analysis we restrict ourselves
to relating only the 2dFGRS spectral classification parameter, $\eta$,
to these processes and leave a fuller development of this subject for
future works.  The investigation presented here is particularly timely
as many new results from the 2dFGRS are being presented which make
exclusive use of this parameter to characterise and partition the
galaxy population (e.g. Madgwick \etal\ 2002; Norberg \etal\ 2002;
Martinez \etal\ 2002).

We make use of two kinds of models to pursue this question. In the
first approach, we adopt a simple parameterisation to describe the
star formation history of each galaxy in terms of an exponentially
decreasing function of time. For a given redshift of observation, the
star formation history is then characterised by a single parameter:
the timescale of the exponential decline. These star formation
histories are then convolved with spectral synthesis models, using the
PEGASE ({\bf P}rojet d'{\bf E}tude des {\bf GA}laxies par {\bf
S}ynthese {\bf E}volutive) code, developed by Fioc \& Rocca-Volmerange
(1997). A particular advantage of this package compared to others
available in the literature (e.g. Bruzual \& Charlot 1993) is that it
is possible to include modelling of nebular emission from star forming
regions in each galaxy --- an important ingredient in the calculation
of the $\eta$ parameter. We refer to these models as the ``simple''
models, because of the rather simplified nature of the parametrised
star formation histories.

The limitation of this simple method is that we must explicitly
assume some family of star-formation histories for our galaxy
population.  We therefore extend this approach to incorporate 
the cosmological framework of the Cold Dark Matter
(CDM) scenario, by using semi-analytic techniques (see \eg\
Kauffmann, White \& Guiderdoni 1993; Cole \etal\ 1994; Somerville \&
Primack 1999).  We make use of a mock catalogue of synthetic spectra
which was designed to match the selection criteria of the 2dFGRS,
analysed previously in Slonim \etal\ (2001), and created using an
updated version of the code described in Somerville \& Primack (1999).
There are several advantages of using a semi-analytic model to create
spectra for this exercise. The star formation history of each galaxy
will reflect at least some of the complexity of the interconnected
processes of dark matter clustering, gas cooling, galaxy merging,
supernova feedback, etc., modelled self-consistently within a specific
cosmological framework. In addition, as we can select galaxies so as
to reproduce the same redshift and luminosity distributions displayed
by the observed 2dF galaxies, the resulting ensemble should contain a
mix of galaxies with different sorts of star formation histories that
is similar to the actual observed sample.  Finally, the semi-analytic
model also yields many other physical parameters for each galaxy,
giving us the potential to further probe the physical processes that
may be related to $\eta$.

We find that in the simple models, there is essentially a one-to-one
relationship between $\eta$ and the ratio of the present rate of star
formation to its past time-averaged value. This quantity is sometimes
known as the ``birthrate'' parameter, $b$ (Scalo 1986). Even more
surprisingly, we find that there is a strong correlation between
$\eta$ and $b$ even in the spectra produced by the semi-analytic
models, despite the complex and diverse nature of the underlying star
formation histories. Empirically, the $b$ parameter is known to be
tightly correlated with morphological type and colour. We therefore
argue that the $\eta$ parameter presents a practical and robust means
of extracting a fundamental measure of galaxy type from the relatively
low-quality spectra typical of present-day large redshift surveys.

Section~2 of this paper briefly reviews the operational definition of
and motivation behind the spectral classification parameter $\eta$,
used in the 2dFGRS.  In Section~3 we make use of the ``simple'' models
based on exponentially declining star formation histories to show how
$\eta$ is related to the star-formation history of a galaxy.
Section~4 then generalises these results using formation histories
generated with a semi-analytic model.  We conclude with a discussion
of our results and ways to build upon the results presented here.

\section{Spectral Classification in the 2dFGRS}
\label{section:section0}

\subsection{Principal Component Analysis}
 
Principal component analysis (PCA) is a well established statistical
technique which has proved very useful in dealing with high
dimensional data sets (see e.g. Murtagh \& Heck 1987; Connolly \etal\
1995).  In the particular case of galaxy spectra we are typically
presented with approximately 1000 spectral channels per galaxy,
however when used in applications this is usually compressed down to
just a few numbers, either by integrating over small line features ---
yielding equivalent widths --- or over broad band filters.  The key
advantage of using PCA in our data compression is that it allows us to
make use of all the information contained in the spectrum in a
statistically unbiased way, i.e. without the use of such ad hoc
filters.
 
In order to perform the PCA on our galaxy spectra we first construct a
representative volume limited sample of the galaxies.  
We compute the ``mean spectrum'' from this ensemble, and subtract this
mean from all of the galaxy spectra.  When we apply the PCA to this
sample an orthogonal set of components (eigenspectra) is constructed,
which span the wavelength space occupied by the galaxy spectra.  These
components have been specifically chosen by the PCA in such a way that
as much information (variance) is contained in the first eigenspectrum
as possible, and that the amount of the remaining information in all
the subsequent eigenspectra is likewise maximised.  Therefore, if the
information contained in the first $n$ eigenspectra is found to be
significantly greater than that in the remaining eigenspectra we can
significantly compress the data set by swapping each galaxy spectrum
(described by 1000 channels) with its projections onto just those $n$
eigenspectra.
 
In the case of the 2dF galaxy spectra we find that approximately two
thirds of the total variance (including the noise) in the spectra can
be represented in terms of only the first two projections ($pc_1$,
$pc_2$).  So, at least to a first approximation, galaxy spectra can be
thought of as a two dimensional sequence in terms of these two
projections.

\begin{figure*}
 \epsfig{file=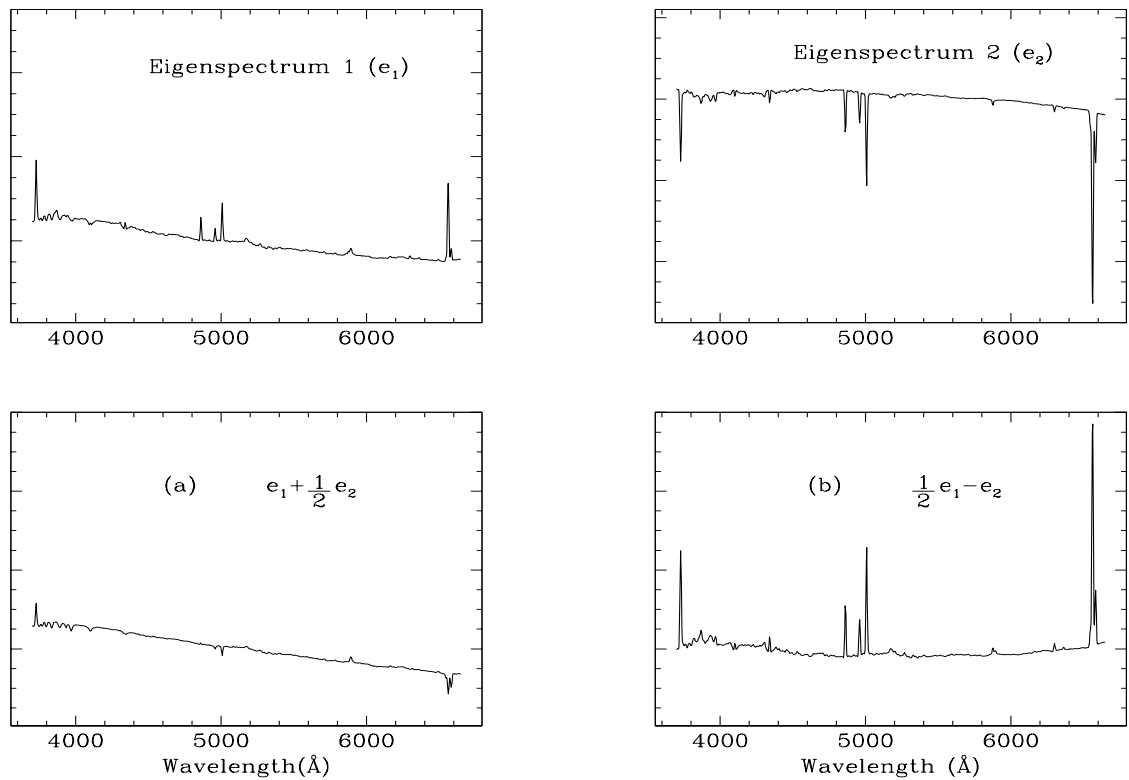,width=7in}
 \caption{The first two principal components identified from the 2dF
 galaxy spectra and the linear combinations which isolate either (a)
 the continuum slope or (b) the line features. Nebular emission
 features are particularly prominent in these spectra, such as [O{\sc{ii}}]
 (3727 \AA) and H$\alpha$ (6563 \AA).}
 \label{fig:cevec}
\end{figure*}

In Fig.~\ref{fig:cevec} we show these first two eigenspectra.  It can
be seen from this figure that whilst the first eigenspectrum contains
information from both lines and the continuum, the second is dominated
by absorption and emission lines.  Because of this it is possible to
take two simple linear combinations which isolate either the continuum
or the emission/absorption line features.  In effect what we are doing
when we utilise these linear combinations is rotating the axes defined
by the PCA to make the interpretation of the components more
straightforward. In so doing we can see that a parameterisation in
terms of $pc_1$ and $pc_2$ is essentially equivalent to a two
dimensional sequence in colour (continuum slope) and the average
emission/absorption line strength.  This ability to isolate the
continuum and line components of the galaxy spectra turns out to be
very useful, as will be discussed in Section~2.2.

\subsection{The $\eta$ Parameter}
\label{sec:eta} 
The 2dF instrument makes use of up to 400 optical fibres with a
diameter of 140$\mu$m (corresponding to $\sim2.0''$ on the sky,
depending on plate position. See Lewis \etal\ 2002).  The instrument
itself was designed to measure large numbers of redshifts in as short
an observing time as possible.  However, in order to optimise the
number of redshifts that can be measured in a given period of time,
compromises had to be made with respect to the spectral quality of the
observations.  Therefore if one wishes to characterise the observed
galaxy population in terms of their spectral properties, care must be
taken in order to ensure that these properties are robust to the
instrumental uncertainties.
 
The quality and representativeness of the observed spectra can be
compromised in several ways and a more detailed discussion of these
issues is presented in previous work (see e.g. Madgwick \etal\ 2002).
The net effect is that the uncertainties introduced into the
fibre-spectra predominantly affect the calibration of the continuum
slope and have relatively little impact on the emission/absorption
line strengths.  For this reason any given galaxy spectrum which is
projected onto the plane defined by ($pc_1$,$pc_2$) will not be
uniquely defined in the direction of varying continuum but will be
robust in the orthogonal direction (which measures the average line
strength).

The linear combination of the first two eigenspectra which is robust
to these uncertainties is shown in Fig.~\ref{fig:cevec}~(b) and
denoted by $\eta$ ({\tt ETA\_TYPE} in the 2dFGRS
catalogue\footnote{\tt http://www.mso.anu.edu/2dFGRS/}). It is simply
\begin{equation}
 \eta = a\cdot pc_1 - pc_2 \;,
\end{equation}
where $a$ is a constant which we find empirically to be $a=0.5\pm
0.1$.

We have now identified the single statistically dominant component of
the galaxy spectra which is robust to the known instrumental
uncertainties.  We have therefore chosen to adopt this (continuous)
variable as our measure of spectral type.  Having defined $\eta$ in
this formal and pragmatic manner, however, we are left with the
question of whether it is physically meaningful, and how it is to be
interpreted.  This is the issue that we now address.

\subsection{Correlation of $\eta$ with traditional galaxy classifiers}
\label{sec:trad}

\begin{figure}
\epsfig{file=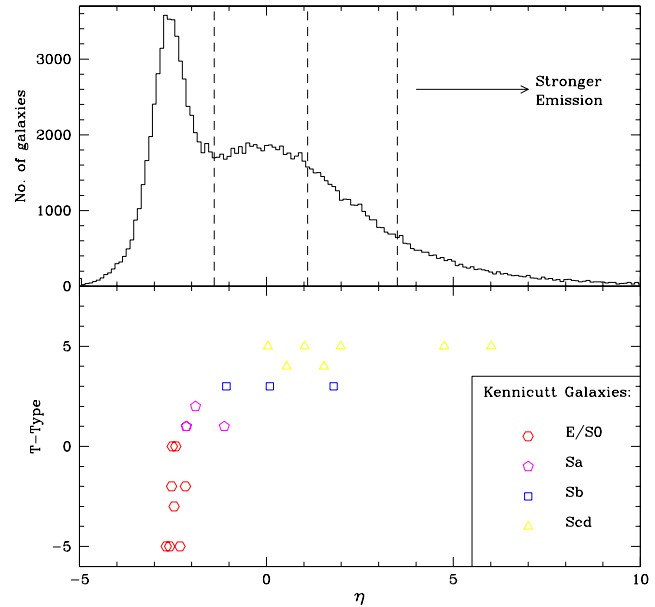,width=3.5in}
\caption{The distribution of $\eta$ projections for the volume-limited
sample of 2dF galaxies (top).  Vertical lines show the divisions in
$\eta$ that correspond roughly to E/S0, early type S, late type S, and
Irr, used in the analysis of Madgwick et al. (2002).  Also shown
(bottom panel) are the $\eta$ projections calculated for a subset of
the Kennicutt Atlas galaxies.  This plot shows a significant
correlation between $\eta$ and galaxy morphology.}
\label{fig:eta}
\end{figure}

We show the observed distribution of $\eta$ for the 2dF galaxies in
Fig.~\ref{fig:eta}. It is rather intriguing that this distribution is
strongly bimodal. Fig.~\ref{fig:eta} also shows the $\eta$ projections
of spectra from galaxies in the Kennicutt Atlas (Kennicutt 1992),
which have known morphologies. There is a correspondence between
morphology and $\eta$ in the expected sense: the value of $\eta$
increases as one moves towards later type objects in the Hubble
sequence. Larger values of $\eta$ indicate more dominant emission
lines, and it is well-known that later type galaxies have stronger
emission lines. From this diagram we see that $\eta$ seems to be
correlated with morphological type, at least based on this small
sample of galaxies. Norberg et al. (2002) showed the distribution of
$\eta$ for a larger sample of galaxies that had been classified into
four morphological types (E, S0, Sp and Irr), and showed that while
there is a correlation between morphological type and $\eta$, there is
a large scatter. The correspondence between $\eta$ and morphological
type is investigated in more detail in Madgwick (2002). 

\begin{figure}
\epsfig{file=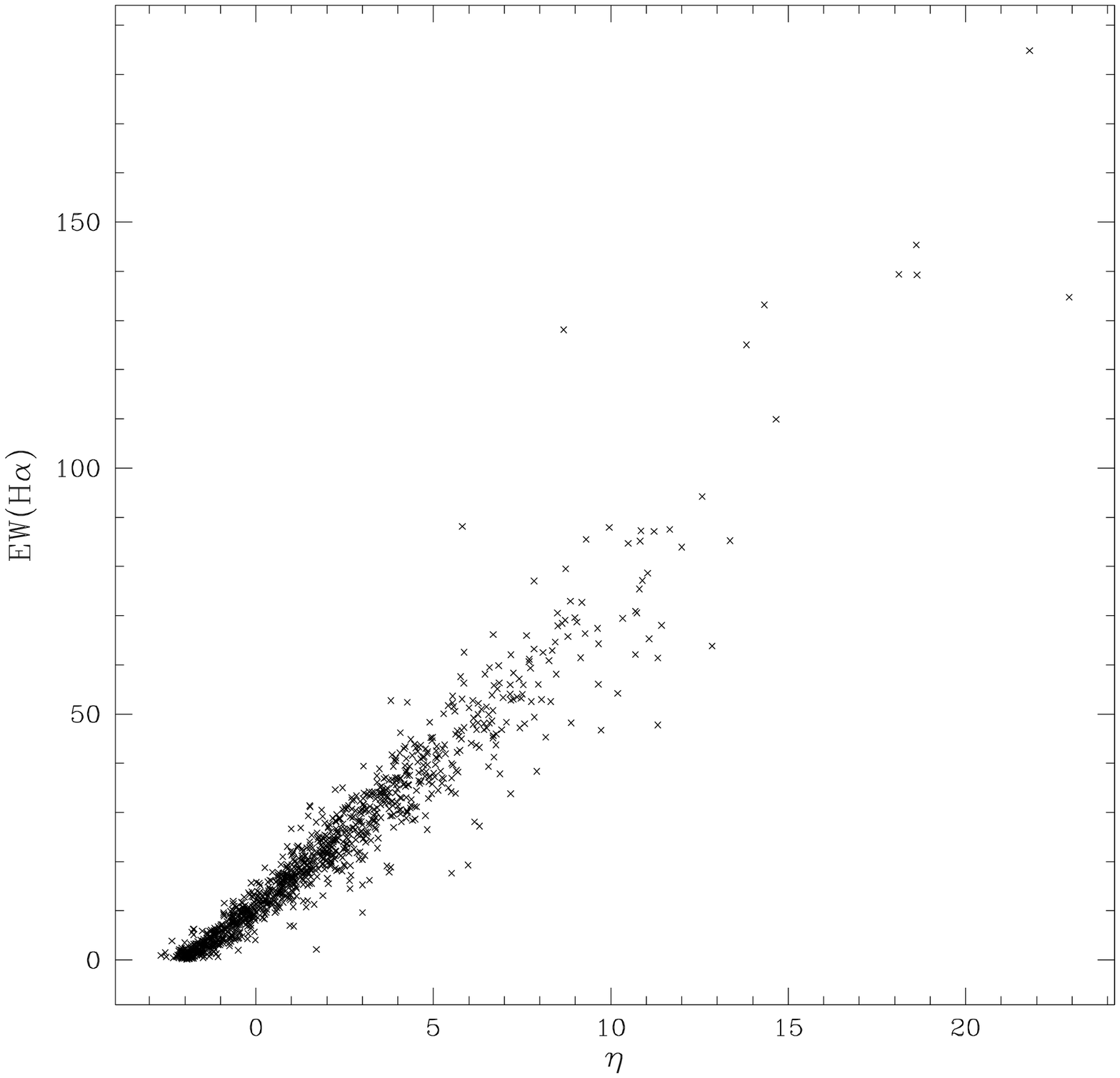,width=3.5in}
\caption{The correlation between the EW(H$\alpha$) for a sample of
high signal-to-noise ratio emission line spectra is shown. The tight
correlation exhibited by these galaxies suggests that $\eta$ is a
measure of relative star formation -- as expressed by the birth-rate $b$
parameter (see Kennicutt, Tamblyn \& Congdon 1994). }
\label{fig:halpha}
\end{figure}

We can also compare the value of $\eta$ derived for each galaxy to the
equivalent width of the H$\alpha$ emission line, for a sample of high
signal-to-noise ratio emission line spectra.  This is shown in
Fig.~\ref{fig:halpha}, from which it can be seen that there is a
strong correlation between these quantities. It is well-known that the
equivalent width of H$\alpha$ is a measure of star formation;
indeed it has been used to estimate the birthrate parameter $b$ (the
ratio of the present rate of star formation to the past averaged value
--- see e.g. Figure~3 of Kennicutt, Tamblyn \& Congdon 1994).  We
return to this point in Section~\ref{section:phys}, where we use
galaxy formation models to investigate how $\eta$ relates to the
star-formation history.

\section{Comparison with Simple Models}
\label{section:section1}

The results shown above suggest that $\eta$ is related both to galaxy
morphology and to star formation history.  In this section we further
explore and test this hypothesis with models produced by combining
simple parameterised star formation histories with spectral synthesis
models. In this way, we can make a direct connection between a
galaxy's star formation history and the value of our spectral
classification statistic, $\eta$. 

\subsection{Model Ingredients}
Stellar population models combine theoretical temperature-luminosity
tracks for stars of various masses with model stellar atmospheres and
an assumed Initial Mass Function (IMF) in order to produce synthetic
spectra for a ``single burst'' population of a uniform age and
metalicity. In order to use these models to obtain predictions for a
composite stellar population (i.e. a galaxy), the ``single burst''
models must be combined with an assumed star formation history, which
provides the distribution of stellar ages and metalicities within the
galaxy. Since the classical work on spectral synthesis by Tinsley
(1980), many authors have adopted a simple parametrisation of the star
formation history in terms of an exponential function of time (e.g. Bruzual
1983):
\begin{equation}
SFR(t) = \frac{1}{\tau} {\rm e}^{-(t-t_{f})/\tau} \;.
\end{equation}
Here, $t_{\rm f}$ is the time at which star formation first
commences, and $\tau$ is the characteristic timescale of star
formation. Most observed spectra can be modelled rather well using
this simple function with the appropriate choice of $\tau$. 
A typical spiral galaxy spectrum might be well-fit with $\tau
\sim 1$ Gyr, while later type galaxies require larger values and early
type galaxies require smaller values of $\tau$.

Nebular emission lines from ionised H$_{\rm II}$ regions and
extinction due to dust also contribute to the appearance of a galaxy
spectrum. Modelling these processes is complicated, requiring
assumptions about the ionization state, the geometry, the metalicity,
the dust mass and the composition of 
H$_{\rm II}$ regions.

The models presented in this Section are based on this simple but
well-defined picture, and are realized using version 2.0 of the PEGASE software
package (Fioc \& Rocca-Volmerange)
In all of the models shown we have adopted a
Kennicutt (1983) initial-mass function with stellar masses in the
range $0.1-120 M_{\sun}$.  In addition, the metalicity of each galaxy
is traced in a self-consistent manner using the
prescription of Woosley \& Weaver (1995), to model the enrichment of
the inter-stellar medium.

Nebular emission lines result from light re-emitted by the ionised gas
in star-forming regions.  A prescription for determining the strengths
of these lines is implemented in the PEGASE code.  This process
involves the absorption of Lyman continuum photons (below 912\AA) by
the nebular gas, which gets ionised, and reaches recombination
equilibrium (Osterbrock 1989).  It is assumed that 70\% of these
photons are absorbed by the gas at solar metalicity. In this approach,
the strength of the nebular emission lines is a function of the age of
the stellar population only, and metalicity and geometric effects are
neglected. 

The PEGASE code provides a simple way to model the effects of dust
extinction on the synthetic spectra, in which the optical depth is
estimated from the mass of gas and the metalicity.  The absorption is
then estimated using observational data for a mixture of graphites and
silicates as in the Milky Way and the Magellanic Clouds (Draine \& Lee
1984 and Pei 1992).  In making this calculation it is also necessary
to make an assumption about the geometry of the galaxy. We
investigated models in which we assumed a spheroidal geometry and also
an inclination-averaged disk geometry.

The spectra generated by the PEGASE code over the optical wavelength
range are generally given in 10\AA\ bins, and the emission line
strengths are specified separately in terms of their peak fluxes over
the continuum.  In order to make comparisons between these spectra and
the 2dFGRS we therefore interpolate the synthetic spectra onto the
4\AA\ binned wavelength range of the 2dFGRS.  Because the given
emission lines are not resolved in this binning we create lines by
superposing Gaussian profiles with the specified peak flux and FWHM
corresponding to that calculated for the 2dF spectrograph using arc
line measurements ($\sim2.3$ pixels). We convert the given synthetic
spectra to units of counts/bin by multiplying the flux (erg/s/\AA) by
the wavelength.  We then normalise each spectrum to have mean counts
of $1$ over our entire wavelength range.  The spectra processed in
this way are directly comparable to those of the 2dFGRS.

\begin{figure*}
\epsfig{file=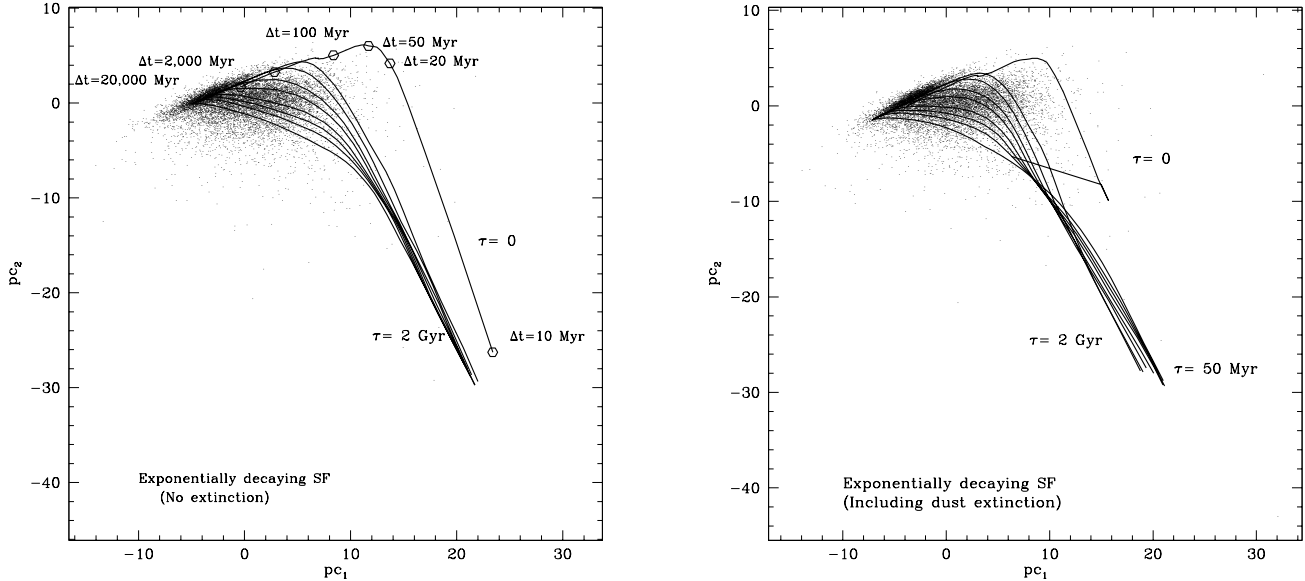,width=7in}
\caption[PCA projections of synthetic galaxy spectra.]  {Dots show the
($pc_1$,$pc_2$) projections for a subset of galaxies in the 2dFGRS
$M_{\bj}-5\log_{10}(h) < -18$ volume-limited sample. Lines show tracks
for model galaxy spectra with simple star formation histories,
parameterised as an exponential function of time. In the left panel,
the model tracks do not include the effects of dust extinction, while
in the right panel, dust extinction (for an inclination-averaged disk 
geometry) 
is included. Each track represents
a different value of the characteristic star formation timescale
$\tau$, while different points along each track represent a different
time since formation, $\Delta t = t-t_{f}$. Labels indicate the
locations along the track of specific values of $\Delta t$.  }
\label{fig:dust}
\end{figure*}

Spectral histories of a given galaxy are compiled for a range of
formation times ($t_{f}=0-20$ Gyr), and a grid of values of $\tau$:
0.05, 0.1, 0.2, 0.3, 0.5, 0.7, 1.0, and 2.0 Gyr. We also calculate
spectral histories for an instantaneous burst of star-formation at
$t=t_{f}$. Because this grid does not comprise a well-defined sample
of galaxies, to compare with the results from the 2dFGRS, we first
subtract the mean spectrum from the 2dFGRS volume limited sample
(described in the previous section) from each spectrum. We then compute
the projection of each spectrum on the eigenspectra derived from the
2dFGRS volume limited sample. The first two projections ($pc_1$,
$pc_2$) derived in this manner are shown for our grid of models in
Fig.~\ref{fig:dust}, along with the corresponding projections for a
random subsample of galaxies in the $M_{\bj}-5\log_{10}(h) < -18$
volume-limited 2dFGRS sample. Model grids are shown both with and
without including the effects of dust extinction. It can be seen that
dust does not have a dramatic impact on the results, but does seem to
improve the agreement between the model tracks and the locus of the
observed galaxies.

Already it is interesting that these simple models seem to cover the
same range of the ($pc1$, $pc2$) parameter space occupied by the
observed galaxy population. Evidently, at least in this simple scheme,
galaxies are ``born'' in the lower right-hand corner of the plot, and
progress towards the upper left-hand corner as they age. As well,
galaxies in the upper right part of the diagram tend to have smaller
values of the characteristic timescale for star formation, $\tau$,
associated with early type galaxies, while as one moves diagonally
towards the lower left corner, galaxies have the more extended
timescales for star formation associated with late type
galaxies. Perhaps unsurprisingly, the vast majority of the 2dF
galaxies (which are fairly luminous) are consistent with times since
formation $\Delta t$ between 2--10 Gyr. It should be kept in mind,
however, that the instrumental effects discussed in
Section~\ref{sec:eta} can introduce random scatter in this diagram. We
therefore turn now to the more robust $\eta$ parameter.

\subsection{The Physical Significance of $\eta$}
\label{section:phys}
We argued based on a visual inspection of the eigenspectrum used to
define the $\eta$ projection that $\eta$ essentially measures the
strength of stellar and interstellar absorption line features and the
strength of nebular emission line features. The strength of the
absorption features mainly indicates the age and metalicity of the
stellar population that dominates the optical luminosity of the
galaxy, and the emission lines are strongly correlated with star
formation activity. The comparison with classical galaxy
classification methods such as visual morphology or equivalent width,
shown in Section~\ref{sec:trad}, suggested that $\eta$ may be
connected with a measure of the star formation \emph{relative} to the
existing older stellar population, such as the birthrate parameter,
$b$ (Scalo 1986). The birthrate parameter is defined as the ratio
between the current and past-averaged star-formation rate, which in
the case of the simple exponential star formation law is given
analytically by
\begin{equation}
\frac{SFR(t)}{\langle SFR(t)\rangle_{\rm past}} =
\frac{(t-t_{f})}{\tau}\frac{{\rm e}^{-(t-t_{f})/\tau}}{(1-{\rm e}^{-(t-t_{f})/\tau})} \;.
\end{equation}
In this section we investigate our hypothesis that $\eta$ might be
correlated with the birthrate parameter, $b$. 

\begin{figure}
\epsfig{file=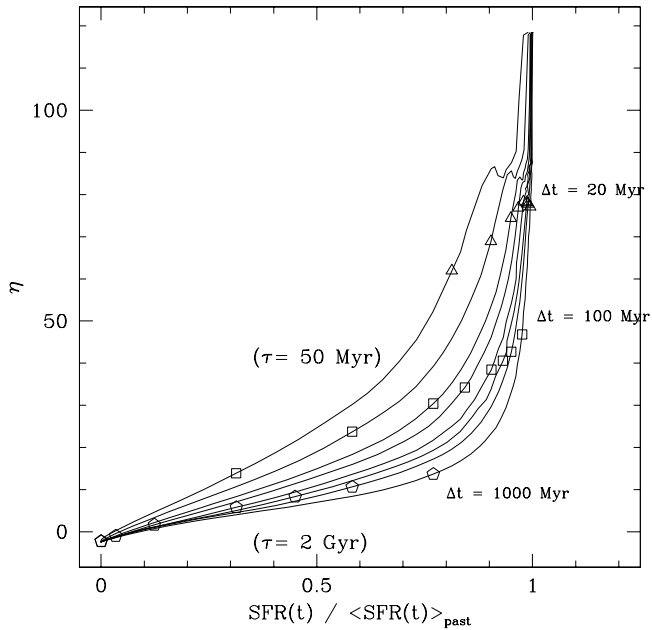,width=3.5in}
 \caption[The ratio of the current star-formation rate to the past
averaged star-formation rate versus $\eta$.]  {The ratio of the
current star-formation rate to the past averaged star-formation rate
(or birthrate parameter, $b$) is plotted against the $\eta$ derived
from the synthetic galaxy spectra for various $\tau$-folding times
(solid lines).  A selection of formation times, $\Delta t = t-t_{f}$,
are also highlighted by the symbols.  It can be seen that for a given
formation time, there is a one-to-one correspondence between $\eta$
and the star formation history as parameterised by the birthrate
parameter $b$. }
\label{fig:stareta}
\end{figure}

The relationship between $\eta$ and $b$ for the grid of models with
simple star formation histories (as before) is shown in
Fig.~\ref{fig:stareta}. For a fixed formation time, $t_f$, there is a
well-defined relationship between the star formation history, as
characterised by the birthrate parameter $b$, and the $\eta$ spectral
parameter. This helps in the interpretation of $\eta$ for the observed
population of galaxies. At least in this simple scheme, galaxies with
small values of $\eta$ formed most of their stars in the past, and are
currently evolving rather passively. This fits in well with the
classical picture of early type galaxy formation. Galaxies with larger
values of $\eta$ have had a significant amount of recent star
formation, characteristic of spiral and irregular galaxies. Recalling
from Fig.~\ref{fig:cevec} that the 2dFGRS sample spans a range of
$\eta$ values of about -5 to 10, this suggests again that most of the
observed galaxies in the 2dF require times since formation greater
than about 1 Gyr. From this diagram, we arrive at a prediction that
the distribution of $\eta$ for galaxies observed at larger look-back
times should exhibit a shift towards higher values. We find that the
effect of extinction corrections is to slightly lower the calculated
$\eta$ for a given star formation scenario, however, this is only a
small effect and does not alter the qualitative interpretation that we
have suggested.

A limitation of the simple characterisation of the star formation
histories that we have adopted here is manifested by the asymptote of
the $b$ values towards unity: because the star formation rate was
assumed to be monotonically decreasing, $b$ is always less than one.
It is clear that we must be cautious in over-interpreting these
results, as real galaxies presumably may have much more complex star
formation histories than we have assumed here. We now pursue a similar
investigation for models with much more complex star formation
histories, generated using a cosmological semi-analytic model of
galaxy formation.

\section{Comparison with Semi-Analytic Models}

In this section, we again extract the $\eta$ parameter from model
galaxy spectra created using stellar population synthesis models. The
difference is that instead of using a simple parameterisation for the
star formation history, the formation history of each galaxy is now
modelled based on physical recipes set within the framework of
hierarchical structure formation. The ensemble of model spectra that
we use here was created using the Somerville et al. semi-analytic
models (e.g. Somerville \& Primack 1999; Somerville, Primack \& Faber
2001) and has been described previously in Slonim et al. (2001). We
refer the reader to those works for details, and here sketch the
ingredients of the models very briefly.

In the hierarchical picture, present-day galaxies such as the ones
observed by the 2dFGRS formed by the merging and accretion of smaller
objects over time. In this picture, the star formation history of a
galaxy is determined by the mass accretion and merging histories of its host
dark matter halo, and the efficiency of gas cooling within those
halos. As the hot gas is enriched with heavy elements by metal-rich
winds from massive stars and supernovae, the cooling efficiency is
increased. Galaxy mergers may trigger powerful bursts of star
formation. In turn, violent star formation events may inhibit future
star formation by heating the interstellar medium or driving winds
that blow it out of the galaxy. 


These star formation histories are convolved with stellar population
models to produce model spectra in much the same way as in the
``simple'' models discussed in the previous section. In the models
used here, we have used the multi-metalicity GISSEL models (Bruzual \&
Charlot, 1983) with a Salpeter IMF to calculate the stellar
part of the spectra. These models are very similar to the PEGASE
stellar population models used in the previous section, and for the
purposes of this investigation, this difference should not
significantly affect our results. 
Nebular
emission lines are added to the spectra in the same way as before,
using the empirical library from PEGASE.

Dust extinction is included using an approach similar to that of
Guiderdoni \& Rocca-Volmerange (1987), which is also very similar to
the approach implemented in the PEGASE package. Here, the mass of dust
is assumed to be proportional to the gas fraction times the metalicity
of the cold gas. We then use a standard Galactic extinction curve and
a ``slab'' model to compute the extinction as a function of wavelength
and inclination. 

The recipes for star formation, supernova feedback, and chemical
evolution involve free parameters, which we set by requiring a halo
with a virial velocity of 220 km/s to host a galaxy with an average
I-band luminosity of about $-21.5+5 \log h_{100}$, and with an average
gas fraction of $0.1$ to $0.2$, consistent with observations of local
spiral galaxies. If we assume that mergers with mass ratios greater
than $\sim$ 1:3 form spheroids, we find that the models produce the
correct morphological mix of spirals, S0s and ellipticals at the
present day. It has previously been shown in numerous works
(e.g. Somerville \& Primack 1999) that this approach leads to fairly
good agreement with numerous key galaxy observations, such as the
local luminosity function, colours, and clustering properties.

We construct a ``mock 2dF catalogue'' of $2611$ model galaxies with
the same magnitude limit, wavelength coverage and spectral resolution,
and redshift range as the 2dF survey.  The synthetic spectra are
expressed in terms of photon counts and the total number of counts in
each spectrum is normalised to unity, as in the prepared observed
spectra.

\begin{figure}
\epsfig{file=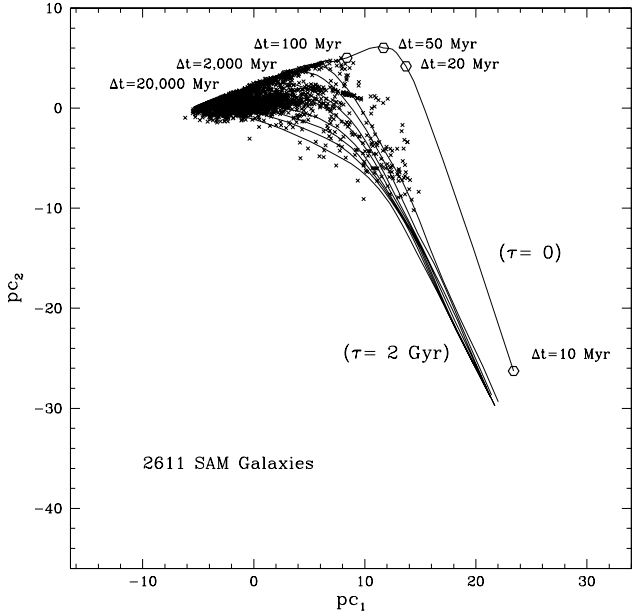,width=3.5in}
\caption{The ($pc_1$,$pc_2$) distribution of the mock galaxy
catalogue.  Also shown are the evolutionary tracks derived from the
``simple'' models of Section~\protect\ref{section:section1}, for a
variety of $\tau$-folding times as before.  It can be seen that there
is a good agreement between the distribution of the mock catalogue
principal components, and those of the observed 2dFGRS galaxies.  The
times highlighted in the figure are the time since formation,
$t-t_{f}$, for the simple model tracks. }
\label{fig:mockstuff}
\end{figure}

The calculation of the strength of the nebular emission features is one
of the less precise aspects of the spectral synthesis packages.  For this
reason in 
Fig.~\ref{fig:mockstuff} we show the first two principal components
($pc_1$,$pc_2$) for the semi-analytic model (SAM) ensemble of synthetic spectra.  It can be
seen that the distribution of the mock catalogue principal components
is quite similar to that of the observed 2dFGRS galaxies.  To guide
the eye, the same set of evolutionary tracks shown in
Fig.~\ref{fig:dust}, derived from the ``simple'' models, is also
shown.  The most noticeable difference between the observed
distribution and that of the mock, is that the former is substantially
broader (most likely due to observational effects such as noise,
reddening and evolution combined with the fact that the observed
galaxies form a {\em much} larger sample). It is also interesting that
the simple model tracks span the same locus of the ($pc1$, $pc2$)
parameter space as the SAM galaxies, despite the fact that as we have
emphasised the star formation histories are very different in the two
kinds of models.

\begin{figure*}
\epsfig{file=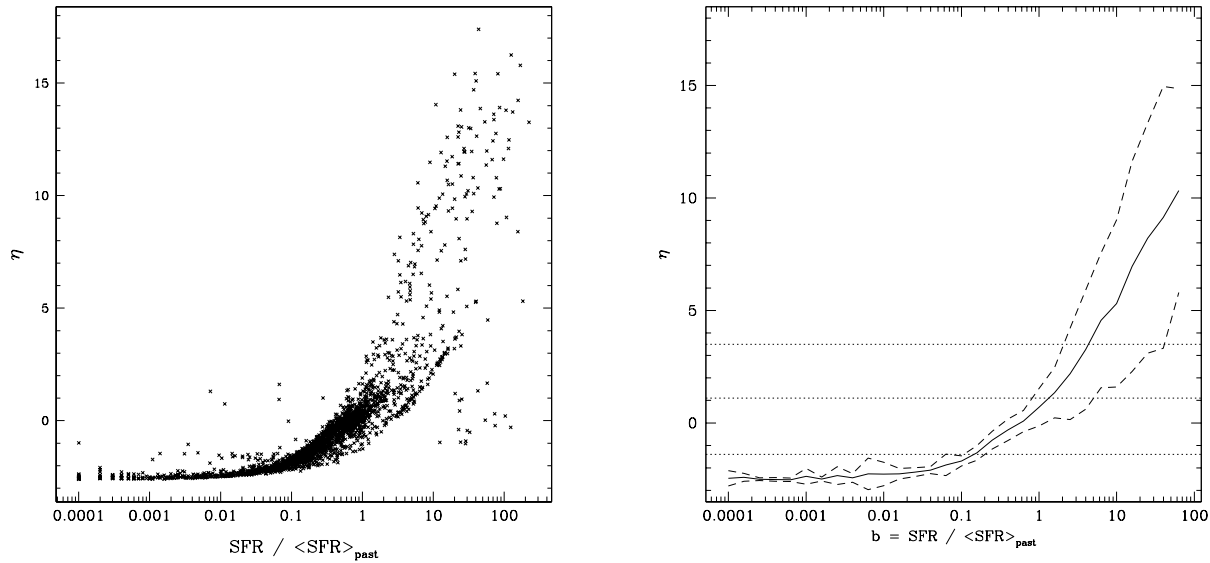,width=6.5in}
\caption[$\eta$ versus the Kennicutt $b$ parameter.]  {The spectral
classification parameter $\eta$ is plotted versus the ratio of the
current to past-averaged star-formation rate (the birthrate, $b$,
parameter) for each galaxy in our SAM mock galaxy catalogue (left
panel).  The right panel shows how the mean value of $\eta$
corresponds to $b$ for these galaxies (the dashed lines show the
$\pm1\sigma$ uncertainties).  Also shown on this plot (dotted lines)
are the cuts in $\eta$ used in Madgwick \etal\ (2002) to calculate the
2dFGRS luminosity functions per spectral type.  It can be seen for
example, that the Type 1 galaxies of that paper ($\eta<-1.4$)
correspond to galaxies with $b<0.1$, or alternatively, galaxies with
current star-formation activity that is only 10\% of their past
averaged rate.}
\label{fig:bkenn}
\end{figure*}

We can now discover whether the strong connection between $\eta$ and
the birthrate parameter $b$ that we demonstrated for the simple models
hold true for galaxies with more arbitrary star formation
histories. In Fig.~\ref{fig:bkenn}, we show $b \equiv SFR/\langle
SFR\rangle_{\rm past}$ vs. $\eta$ for the mock-2dF ensemble of SAM
galaxy spectra. Again we see a strong relationship between $\eta$ and
the $b$ parameter. Also, it is encouraging to note that the range of
$\eta$ spanned by the SAM galaxies (-5 to 15) is similar to that
spanned by the 2dF galaxies. The fact that the spectral properties of
the galaxies in the SAM ensemble appear similar to those of the real
2dF galaxies enables us to use the results of the SAMs to draw a
correspondence between the actual numerical value of $b$ and
$\eta$. This leads to the interesting conclusion that the dividing
line between the two ``bumps'' in the bimodal distribution of $\eta$
seen in Fig.~\ref{fig:cevec}, $\eta \sim -1.4$, corresponds to
galaxies that are forming stars at about 1/10th of their past-averaged
rate. Similarly, the other two $\eta$ thresholds adopted by Madgwick
et al. (2002) to calculate luminosity functions per spectral type may
be matched up with $b$ values using Fig.~\ref{fig:bkenn}.

We have also investigated correlations between $\eta$ and other
physical properties of galaxies in the SAMs, such as bulge-to-total
ratio or mean stellar age. While other correlations exist, none are as
tight as the correlation shown in Fig.~\ref{fig:bkenn} between $\eta$
and the birthrate parameter $b$. Of all the correlations we
investigated, the connection between $\eta$ and star formation history
seems to present the most straightforward interpretation of the $\eta$
parameter. We defer a more detailed investigation of the other
physical correlations in the SAMs and comparison with the 2dF and
other data sets to a future work.

\section{Discussion}
\label{section:conclusion}

In this paper we have studied the physical interpretation of a
PCA-based spectral parameter, $\eta$, defined for the analysis of the
2dFGRS spectra (Madgwick et al. 2002).  Although the definition of
$\eta$ was motivated by the need to compromise between the desire to
extract the maximum amount of statistical information from the spectra
(in the sense of PCA's maximum variance) and the limitations imposed
by the instrumental uncertainties of the 2dF, we have argued here that
this parameter has a straightforward and physically meaningful
interpretation. We find that $\eta$ is an indicator of the star
formation history of the galaxy, and is tightly correlated with the
birthrate parameter $b = SFR/\langle SFR \rangle_{\rm past}$, which
characterises the ratio of present to past-averaged star formation.

A first indication of this correspondence is the correlation we find
between $\eta$ and and the equivalent width of the H$\alpha$ emission
line, EW(H$\alpha$), which has been used as a
direct measure of the birthrate parameter in previous studies
(e.g. Kennicutt, Tamblyn \&
Congdon 1994). We show that in models with simple star formation
histories, parameterised in terms of an exponential function of time,
there is a one-to-one correspondence
between $\eta$ and the birthrate parameter $b$ (for a fixed formation time). 
Perhaps this is not
too surprising in the context of classical spectral synthesis work ---
it has long been known (e.g. Tinsley 1980) that a simple SF history
can be used to interpret spectral appearance, i.e. $\eta$ in this
context. More surprisingly, we show also that a strong relationship
between $\eta$ and $b$ is also exhibited by model galaxies with much
more complex star formation histories, created within a hierarchical
cosmological framework using a semi-analytic model. 

The explanation of this surprisingly simple result must lie in the physical
processes that determine the appearance of a galaxy spectrum. First,
it is important to remember that at optical wavelengths, where all of
this analysis has been carried out, the spectrum is dominated by the
most luminous stars, and is therefore biased towards the most recent
significant star formation activity.  The slope of the optical
continuum and the strength of stellar absorption lines evolve on the
timescale of the lifetimes of intermediate type main sequence stars,
roughly several Gigayears. The strengths of nebular emission features,
however, evolve on much shorter timescales, as they require the
presence of very hot, short lived O and B type stars. These features
therefore depend on the star formation history on the timescale of
5--10 Megayears. The $\eta$ parameter has been defined so as to be
insensitive to the continuum slope (for practical reasons), and 
hence represents a sequence from
strong absorption lines to strong emission lines.  It is well-known
that the strengths of stellar absorption lines (especially Hydrogen
recombination lines such as the Balmer series) are good indicators of
stellar age, while the strength of the nebular emission lines are
indicators of present star formation rate. With all this in mind, we
can deduce that the rapid evolution in the value of $\eta$ over the
first several hundred Myr of a galaxy's life (see
Fig.~\ref{fig:stareta}) reflects the fading of the emission lines as
the star formation rate declines and the young stars burn
out. Galaxies then tend to ``pile up'' at low values of $\eta$ as the
age of their dominant stellar population exceeds a few Gyr.

In this sense it is straightforward to understand the relationship 
between $\eta$ and $b$ for our ``simple'' models, of
{\em monotonically} decreasing star formation activity.
However,
the star formation histories of galaxies created by the semi-analytic
models can be quite complex. The star formation rate fluctuates
dramatically and non-monotonically over time as the galaxy exhausts
its gas and then accretes a new gas supply, or experiences bursts of
star formation triggered by mergers.  The issue we must now address is how
an optical diagnostic such as $\eta$, which is dominated by only the most
recent events of star-formation, can be so closely related to the birth-rate 
parameter, $b$, which incorporates the entire formation history of a galaxy.

The basis of the hierarchical picture of galaxy
formation --- from which our mock catalogue of SAM galaxies has been derived ---
is that galaxy properties are determined by the merger
histories of their host dark matter halos. 
In essence, the history of a specific halo (or galaxy) can be
understood in terms of how the 
density on the scale of that structure compares to the
background density.  A region with density much above the average (a
``many sigma'' peak, in the language of Gaussian random fields) 
collapses early, while lower density peaks collapse
later. It can therefore be conjectured that the early collapse
of a dense dark matter halo can be associated with early star formation, early
consumption of all available gas, and low present-day star formation
rates. Conversely, late-forming objects will have a sustained gas
supply and ongoing star formation. This suggests that the birthrate
parameter $b$ reflects perhaps the most theoretically fundamental
property of a galaxy (or of its host halo): small-$b$ galaxies
represent rare, many-sigma peaks in the primordial density field,
while large-$b$ galaxies are formed in more common, lower density
peaks.  This trend is manifested in the numerous empirical
correlations between galaxy ``type'' (as characterised by morphology,
colour, or spectral type), luminosity, and environment. 
We expect this simple picture
to be complicated by the details of gas cooling, star formation,
feedback, etc., but the results from the semi-analytic models (which
include all of these effects at some level, although of course the
real Universe is likely to be even more complicated) suggest that this
only introduces a moderate amount of scatter on top of the general
trend.

We therefore conclude that the $\eta$ parameter represents a promising
candidate for galaxy classification in large modern redshift
surveys. It was adopted in previous analyses of the 2dFGRS because of
its practical advantages: it is straightforward and efficient to
compute in an automated fashion, and it is robust to the instrumental
uncertainties commonly associated with fibre-based multi-object
spectrographs. Here we have shown that $\eta$ also has a
straightforward physical interpretation which can be intuitively
connected both with traditional classifiers such as morphology, and
that it seems to be connected with theoretically fundamental
properties of galaxies within the modern hierarchical structure
formation paradigm.

\section*{Acknowledgments}

The efforts of the 2dFGRS collaboration, in preparing and compiling
the data used in this analysis, are greatly appreciated.  In particular
we thank Karl Glazebrook for providing many useful comments and suggestions.
DSM was
supported by an Isaac Newton studentship from the Institute of
Astronomy and Trinity College, Cambridge. RSS thanks the IoA for
hospitality during completion of this work, and acknowledges support
from the IoA visitor program.

\bsp
\label{lastpage}
\end{document}